\documentclass[iop,numberedappendix]{emulateapj}
\usepackage{graphicx}
\usepackage{bm}
\usepackage{amsmath}
\usepackage{natbib}
\usepackage{color}
\usepackage{multirow}

\newcommand{\eqb}{\begin{eqnarray}}
\newcommand{\eqe}{\end{eqnarray}}
\newcommand{\diff}{\textrm{d}}

\newcommand{\pperp}{p_{\perp}}

\newcommand{\ppar}{p_{\parallel}}
\newcommand{\pparb}{p_{\parallel 0}}

\newcommand{\gammaw}{\gamma_{\rm w}}

\newcommand{\betaw}{\beta_{\rm w}}
\newcommand{\nuw}{\nu_{\rm w}}

\newcommand{\av}[1]{\left\langle #1 \right\rangle}

\newcommand{\derdrho}{\frac{1}{\rho^2}\frac{d}{d \rho}}

\newcommand{\der}[1]{\frac{\partial}{\partial #1}}

\newcommand{\derd}[1]{\frac{d}{{d} #1}}

\newcommand{\pparexpand[1]}{{p_\|^{(#1)}}}

\newcommand{\gammaexpand[1]}{{\gamma^{(#1)}}}
\newcommand{\nexpand[1]}{{n^{(#1)}}}
\newcommand{\Deltaexpand[1]}{{\Delta^{(#1)}}}
\newcommand{\pperpexpand[1]}{{p_\bot^{(#1)}}}
\newcommand{\Eexpand[1]}{{E^{(#1)}}}
\newcommand{\Bexpand[1]}{{B^{(#1)}}}

\newcommand{\deltaexpand[1]}{{\delta^{(#1)}}}

\newcommand{\tableone}{
\begin{deluxetable*}{ccccccc}
\tablecaption{\label{pwntable}%
Pulsar wind nebulae studied by \citet{2011MNRAS.410..381B}.}
\tablecolumns{7}
\tablehead{
\colhead{} & \colhead{Crab} & \colhead{3C~58} & \colhead{B1509-58} & \colhead{Kes75} & \colhead{W44} & \colhead{K2/K3 Kookaburra}}
\startdata
References		   & 1 & 2, 3 & 4,5 & 6 & 7, 8 & 9 \\
$P_{\rm pulsar}$ (ms) 		   & 33 & 65 & 150 &  324.8 & 267 & 68.2 \\
$L (\times10^{38}$ erg s$^{-1}$)  & 5   & 0.27 & 0.18 & 0.083 & 0.0043 & 0.1 \\
$a_{\rm L}(\times10^{10})$ & 7.6 & 1.8  & 1.4  & 0.98  & 0.22   & 1.1  \\
$\kappa$ (lower limit)       & $10^6$ & $5\times10^5$ & $3\times10^5$ & $10^5$ & $10^5$ & $10^5$ \\
$\mu$ (upper limit)          & $1.9\times10^4$ & $8.8\times10^3$  & $1.2\times10^4$ & $2.4\times10^4$ & $5.5\times10^3$ & $2.7\times10^4$ \\
$P_{\rm conv}$ (dyn cm$^{-2}$)   & $\sim10^{-8}$ & $3.2\times10^{-10}$ & $\sim10^{-10}$ & $\sim10^{-11}$ & $>5.7\times10^{-10}$ & $4\times10^{-10}$ \\
Conversion radius  & \multirow{2}{*}{575}   & \multirow{2}{*}{781} & \multirow{2}{*}{800} & \multirow{2}{*}{2300} & \multirow{2}{*}{86} & \multirow{2}{*}{1979} \\ 
$R_{\rm conv}$ (upper limit) & & & & & &
\enddata
\tablerefs{(1) \citealt{1984ApJ...283..694K}; (2) \citealt{2002ApJ...568..226M}; (3) \citealt{2002ApJ...571L..45S}; (4) \citealt{1999MNRAS.305..724G}; (5) \citealt{2009PASJ...61..129Y}; (6) \citealt{2008ApJ...686..508N}; (7) \citealt{1996ApJ...464L.165F}; (8) \citealt{1999ApJ...524..179C}; (9) \citealt{2005ApJ...627..904N}}
\end{deluxetable*}
}

\slugcomment{Submitted to ApJ}
\shorttitle{Superluminal Waves}
\begin{document}

\title{Propagation and stability of superluminal waves in 
pulsar winds}
\author{Iwona Mochol and John G. Kirk}
\affil{Max-Planck-Institut f\"ur Kernphysik, Postfach 10~39~80,
69029 Heidelberg, Germany}
\email{iwona.mochol@mpi-hd.mpg.de, john.kirk@mpi-hd.mpg.de}

\begin{abstract}
 
Nonlinear electromagnetic waves with superluminal phase velocity can propagate in the winds around isolated pulsars, and around some pulsars in binary systems. Using a short-wavelength approximation, we find and analyze an integrable system of equations that govern their evolution in spherical geometry. A confined mode is identified that stagnates to finite pressure at large radius and can form a precursor to the termination shock. Using a simplified criterion, we find this mode is stable for most isolated pulsars, but may be unstable if the external pressure is high, such as in the pulsar wind nebulae in starburst galaxies and in W44.  Pulsar winds in eccentric binary systems, such as PSR~1259-63, may go through phases with stable and unstable electromagnetic precursors, as well as phases in which the density is too high for these modes to propagate.  

\end{abstract}


\section{Introduction}

Electromagnetic fields, modulated at the rotation frequency of the
neutron star, form the energetically dominant component of pulsar
winds.  These flows are responsible for transporting the rotational
energy lost by the star and depositing it in the surrounding pulsar
wind nebula (PWN). As well as energy, they also convey the magnetic
flux and the charged particles --- most likely electrons and positrons
--- that are required to produce synchrotron radiation in the PWN 
\citep{1974MNRAS.167....1R}.

An important property that sets pulsar winds apart from other stellar winds 
is their relatively low density. As a result, the fluctuations 
imposed by the rotation of the neutron star are able
to propagate not only as MHD waves frozen in to the outflowing plasma, 
but, beyond a critical or cut-off radius $r_{\rm c}$, 
also as large-amplitude electromagnetic modes of superluminal phase
velocity 
\citep{1975A&A....44...31A,1996MNRAS.279.1168M}.  
The location of the
cut-off radius depends on the wave amplitude and on the relative
strength of the phase-averaged fields and the fluctuating components,
which, in turn, depend on the obliquity of the pulsar (i.e., the 
angle between its magnetic and 
rotation axes) and on latitude in the wind. 
For isolated pulsars it generally lies well
inside the position where a termination shock can be expected 
\citep{2012ApJ...745..108A}, so that superluminal waves may play an
important role in the outer parts of the wind and also in the
termination shock itself. In particular, since they may carry 
the entire wind luminosity and are notoriously unstable 
\citep{1973PhFl...16.1480M,1974PhFl...17..778D,1980PhRvA..22.1293A,1980JPlPh..24...89L},
they could provide the key to 
resolving the well-known \lq\lq $\sigma$-problem\rq\rq. 

This term is used to describe the lack of a convincing mechanism for
converting Poynting flux into particle-carried energy flux. Poynting
flux dominates the wind at launch, but is thought to be a small
fraction of the energy budget outside the termination shock.  Recent
work 
\citep{2012arXiv1212.1382P} 
suggests that MHD instabilities in the shocked wind may,
as proposed by 
\citet{1998ApJ...493..291B},
be able to reduce $\sigma$ (the ratio of Poynting flux to kinetic-energy flux) 
from a starting value of a few just outside the shock, to the value suggested by
observation, which is $\sim10^{-3}$ for the Crab Nebula 
\citep{1984ApJ...283..710K}.

However, the mechanism underlying the transition from $\sigma\gg1$
near the pulsar to $\sigma\sim1$ at the termination shock remains
controversial. Most work on this problem has concentrated on the
damping of MHD like wave-modes either in the wind 
\citep{2001ApJ...547..437L, 2003ApJ...591..366K, 2010ApJ...725L.234L}  
or at the termination shock itself, where current sheets carried into
the shock are compressed, giving rise to dissipation by driven reconnection. 
\citep{2003MNRAS.345..153L, 2008ApJ...682.1436L, 2007A&A...473..683P}. 
In this scenario, particle acceleration is primarily associated with
the dynamics of the reconnection region  
\citep{2011ApJ...741...39S}.

However, it has recently emerged that the termination shock
structure may be very different in the region where superluminal modes
can propagate.  In this case, the mechanism of dissipation is closely
connected with the parametric instabilities of these modes, rather
than with driven reconnection 
\citep{amanokirk13}.  
The associated particle acceleration mechanisms have not yet been
investigated in detail, but it is suggested that the
first order Fermi process may be much more prominent 
when superluminal waves are present, 
than in the case
of dissipation by reconnection.

This question is central to the study of PWN, and especially their high energy (TeV) emission. 
Current models adopt ad~hoc assumptions concerning the injection of accelerated particles
at the termination shock 
which take no account of the role of the surrounding medium in determining the shock
location and structure. 
In order to establish a predictive theory of pulsar winds, it is, therefore, 
fundamentally important to identify which PWN 
are likely to sustain shocks mediated by superluminal modes, 
and which are not, and to classify important properties such as the stability or 
instability of these modes at the position of the shock. 
These are the questions we address. Our approach builds on the work of 
\citet{2012ApJ...745..108A}, 
  who found the shape of the cut-off surface $r=r_{\rm c}$ for both
  linearly and circularly polarized waves of arbitrary amplitude, 
including, for the former, a
  non-zero value of the phase-averaged magnetic field.  
Here, we develop a perturbation method to determine 
how a superluminal mode launched outside the cut-off surface
will evolve as it propagates radially outwards, searching 
for those regions of the wind in which such a packet remains
relatively stable, and those in which it can be expected to thermalize rapidly.

In section~\ref{largeamplitudewaves} we recall the description of
nonlinear superluminal plane waves in a two-fluid electron-positron
plasma, and 
summarize the literature concerning their stability
properties. Radially propagating waves in spherical geometry are treated in 
in section~\ref{sphericalwaves}. First, the 
short-wavelength perturbation theory is developed and is shown to
result in an integrable system of equations 
for the radial evolution of wave packets, consisting of  
the conservation laws of particles and energy, supplemented by an analogue 
of the 
entropy equation. Restricting, for simplicity, 
the treatment to circularly polarized modes, it is shown
that two kinds of wavepacket are possible --- freely expanding
modes, which ultimately turn into vacuum waves,
and confined modes, which stagnate at finite pressure, and can be identified
as extended precursors of the termination shock.
Section~\ref{stabilityzones} discusses the confined modes and shows that 
two distinct regions exist in which they are relatively stable. One 
of these is relevant for pulsar winds in high pressure environments, such as 
the high-pressure 
wind of a companion star, the other is relevant for isolated pulsars
in low-pressure surroundings, such as supernova ejecta. 
Analytical constraints are given 
on the values of pulsar period, surrounding pressure
and mass-loading parameter $\mu$ that determine whether or not  
a pulsar wind has the potential 
to generate a stable superluminal precursor at its termination shock. 
In section~\ref{discussion} 
we discuss the significance and limitations of our results and their 
implications for PWN around isolated pulsars and in gamma-ray binaries.
Finally, a brief summary of our main results is presented in section~\ref{conclusions}. 

\section{Large-amplitude plane waves}
\label{largeamplitudewaves}

\subsection{Basic properties}

The simplest system able to describe superluminal modes is that of two cold
charged fluids. It has been analyzed in detail by  
\citet{clemmow74,clemmow77}.
In the pulsar case, we are concerned with cold electron and positron fluids
and, in particular, with
electromagnetic modes in which the displacement current
is non-zero in all frames of reference. 

In the outer parts of a pulsar wind, the radial component of the magnetic 
field is very small, and will be neglected in the following. 
In this case, radially propagating electromagnetic
waves have only transverse fields, but 
the fluid velocities can have a component in the direction
of propagation. The waves 
are characterized by a 
superluminal phase speed $\beta_{\rm phase}>1$ and a  
subluminal group
speed $\betaw=1/\beta_{\rm phase}$ and will henceforth be called simply 
\lq\lq superluminal\rq\rq\ modes. 
The electron and positron fluids have the same 
proper number density $n$, and move with the
same velocity $\ppar/\gamma$  (in units of $c$) 
in the direction of wave motion. But the
transverse velocity of the positrons $\vec{p}_\perp/\gamma$ 
is antiparallel to that of the electrons. (Here, $\gamma$ is the Lorentz 
factor of the fluids, and $\ppar$ and $\vec{p}_\perp$ are components of the 
dimensionless four-velocity of the positron fluid). 
This generates a transverse conduction current.
In the homogeneous or \lq\lq H-frame\rq\rq, in which the group speed vanishes, 
the fluid and field variables are
space-independent, and the conduction current exactly balances the 
displacement current.  
In general, $\ppar/\gamma$ differs from the wave group speed
$\betaw$, implying a non-vanishing flux of particles in the 
H-frame.

The governing equations have been presented elsewhere 
\citep[in][for example]{2012ApJ...745..108A},
and are reproduced in Appendix~\ref{app1}. The 
nonlinear solutions take their simplest form in the H-frame, 
where the phase variable (\ref{wkbphase}) is purely temporal. 
Circularly polarized modes are monochromatic (i.e., depend
sinusoidally on phase)
and have phase-independent $\ppar$. The phase-averaged fields vanish. 
Linearly polarized modes, on the other hand, 
have a \lq\lq saw-tooth\rq\rq\ dependence of 
$\ppar$ on phase, and may also carry 
a non-zero phase-averaged transverse
magnetic field component perpendicular to the 
oscillating electric field.

Although the analytical treatment of linearly polarized modes is significantly
more cumbersome, their dispersion curves, at least for vanishing
phase-averaged magnetic field, are very similar to those of 
circularly polarized modes 
\citep{2012ApJ...745..108A}. 
Consequently, we restrict much of 
the following discussion to circularly polarized modes.

\subsection{Stability}

In unmagnetized plasmas, strong electromagnetic waves are subject to
parametric
instabilities. These are induced by longitudinal density fluctuations 
which can couple
to the transverse electromagnetic side-band modes 
and cause backscattering, filamentation, or absorption of
the driver \citep{1973PhFl...16.1480M,1974PhFl...17..778D}. The growth
rates can be as high as the wave frequency.  The stability of
self-consistent waves was first investigated in a
nonrelativistic electron-ion plasma
\citep{1972PhRvL..29.1731M,1973PhFl...16.1480M}. This work
was extended to the relativistic case 
for electron-positron plasmas by \citet{1978JPlPh..20..479R} and
\citet{1978JPlPh..20..313L}. In all cases only perturbations
that propagate in the direction of driver's motion were discussed, and
the dispersion relations were obtained by linearization.  In general, 
it was found that both short- and long-wavelength perturbations
are unstable. Finite-temperature effects
can be expected to suppress the instabilities at short wavelengths,
although, to our knowledge, a complete analysis is lacking.
Long wavelength
perturbations, on the other hand, are stable if the 
particle flux through the wave in the H-frame is sufficiently
large. A simple criterion 
is given by \citet{1978JPlPh..20..313L}. Denoting quantities measured in the 
H-frame by a prime, they find stability 
when
\eqb 
S&\equiv&\ppar'^2-2\gamma'\pperp+\pperp^2\,>\,0\enspace.
\label{stcond}
\eqe 
Numerical studies of \citet{1978JPlPh..20..479R} and the PIC
simulations of \citet{2005ApJ...634..542S} confirm the stabilizing
effect of relativistic streaming. Although no precise test
of it has been undertaken, we nevertheless 
adopt (\ref{stcond}) in the following as the stability criterion 
for circularly polarized 
modes, noting that for linearly polarized modes
the presence of a phase-averaged transverse component of
the magnetic field has an additional stabilizing effect 
\citep{1980PhRvA..22.1293A}. 

\subsection{Electromagnetic Hugoniot curve}
\label{hugoniot}

Plane waves in this model can be characterized by the phase-averaged values 
of their particle flux
density $J$, energy flux per particle $\mu$, and 
parallel (i.e., in the propagation direction) momentum flux per particle
$\nu$.  In a local analysis, waves in a radial pulsar wind 
can also be
considered plane. A radial wind occupying a total solid angle 
$\Omega_{\rm s}$ is characterized 
by an energy flux $L$ 
and a flux of electrons and positrons of $\dot{N}$, in terms of which
the parameters $\mu$ and $J$ are 
\eqb
\mu&=&L/\left(\dot{N}mc^2\right) \enspace,
\\
J&=&\dot{N}/\left(\Omega_{\rm s}r^2\right)\enspace.
\eqe
The momentum flux is connected with the magnetization parameter $\sigma$. 
In a cold, striped wind, it is 
\citep{2010PPCF...52l4029K}
\eqb
\nu&\approx&\mu-\frac{\sigma}{2\mu}+\frac{\sigma^4}{8\mu^3}
\label{approxnu}
\eqe
(for $\mu\gg1$
and $\sigma\lesssim\mu^{2/3}$)
and is independent of radius, 
provided there is no dissipation.
It is related to the ram pressure $P$ (the $(r,r)$
component of the stress-energy tensor) by
\eqb
P&=&\frac{\nu}{\mu}\left(\frac{L}{\Omega_{\rm s}r^2 c}\right)\enspace.
\eqe
Observations of the Crab Nebula suggest (e.g., \citealt{2011MNRAS.410..381B})
\eqb
L&=&5\times10^{38} \textrm{ erg s}^{-1} \enspace,
\\
\dot{N}&=&10^{40} \textrm{ s}^{-1}\enspace,
\eqe
so that 
$\mu\approx10^4$. Close to the star, a magnetically dominated 
wind is expected to accelerate 
rapidly until mildly super(magneto)sonic 
\citep{2009ASSL..357..421K}
implying $\sigma\lesssim\mu^{2/3}$, so that 
$\nu\approx\mu$.
For nonlinear waves, the wave group speed depends not only on the
frequency, but also on other wave properties, such as the
amplitude. Thus, in order to plot a dispersion curve, giving, for
example the group speed as a function of frequency, additional
constraints are needed. Choosing $\mu$ and $\nu$ to be constant along
such a curve transforms it into the electromagnetic equivalent of a
Hugoniot curve: it then specifies all the (plane) 
waves of a given frequency that can
be launched in a pulsar wind characterized by the same values of 
$\mu$ and $\nu$ 
\citep{2012ApJ...745..108A}.

\section{Spherical waves}
\label{sphericalwaves}

\subsection{Short-wavelength approximation}

The radial evolution of spherical nonlinear waves 
at distances from the origin large compared to the wavelength
can be treated using standard perturbation
techniques 
\citep{1984A&A...139..417A,2011ApJ...729..104K,2011ApJ...736..165K}.
These lead straightforwardly to equations that express conservation
of the phase-averaged particle and radial energy fluxes (Appendix~\ref{app1})
\eqb
\frac{1}{r^2}
\frac{\partial}{\partial r}\!
\left(r^2\! \left<2n\ppar\right>\right)&=&0\enspace,
\label{continuity1}
\\
\frac{1}{r^2}\frac{\partial}{\partial r}\!\left(
r^2\! \left<2n\ppar
\gamma\right>
+\frac{r^2\betaw\left<\!
\left|E\right|^2\!\right>}{4\pi mc^2}\right)
&=&0\enspace.
\label{energy1}
\eqe
However, the divergence of the $(r,r)$ component of the stress-energy tensor
does not yield a conservation law directly:
\eqb
\frac{1}{r^2}\frac{\partial}{\partial r}\!
\left(
r^2\! \left<2n\ppar^2\right>
+\frac{r^2\!\left(1+\betaw^2\right)\!
\left<\! \left|E\right|^2\!\right>}{8\pi m c^2}\right)
&=&\frac{\left<n\pperp^2\right>}{r}\enspace.
\label{momentum1}
\eqe 
Nevertheless, as we show in the Appendix~\ref{appendixentropy},
an integral of motion can be found by constructing the electrodynamic
equivalent of the hydrodynamic entropy equation, essentially by
subtracting $\betaw$ times the energy equation (\ref{energy1}) from
the momentum equation (\ref{momentum1}).  
This integral can be found
  more directly by considering the adiabatic invariants of the
  zeroth-order plane wave.\footnote{We are indebted to an anonymous
    referee for suggesting this approach.} 
In both the lab.\ (pulsar)
frame and the H-frame, the charge density of the wave vanishes, and
both the current and the electric field are 
entirely transverse. This means that 
the electrostatic potential $A^0$ vanishes when the Coulomb gauge is chosen.
It is then a trivial matter to construct an adiabatic invariant $\Phi$  
of the motion of an individual fluid particle 
by integrating over one period of oscillation the zeroth
component of the canonical momentum $P^\mu$:
\eqb
\Phi&=&\oint \diff t\, P^0
\nonumber\\
&=& \oint \diff t\, \left(mc\gamma +e A^0/c\right)
\nonumber\\
&=&2\pi mc\left<\gamma\right>/\omega
\eqe
where $\omega$ is the angular wave frequency in the lab.\ frame.
In our case, $\omega$ 
equals the (constant) angular velocity of the pulsar, so that  
the invariance of $\Phi$ in a slowly expanding (or contracting) flow 
implies
\eqb
\frac{\partial}{\partial r}\left<\gamma\right>&=&0\enspace.
\label{entropy1}
\eqe
In addition, to these equations, Amp\`ere's law 
provides a connection between
the electric field and the fluid momentum:
\eqb
\frac{\partial \vec{E}}{\partial t}&=&-mc
\omega_{\rm p}^2\gammaw^2\vec{p}_\perp/e\enspace,
\label{ampere1}
\eqe
where 
$\gammaw=\left(1-\betaw^2\right)^{-1/2}$ is the Lorentz factor
associated with the wave group speed, and 
$\omega_{\rm p}=\left(8\pi n e^2/m\right)^{1/2}$ is the proper plasma frequency. 

\subsection{Circular polarization}
\label{circular}

For simplicity, we now restrict the treatment to circular polarization, 
the corresponding expressions for linear polarization are 
given in Appendix~\ref{linear}. 

To close the system, a non-linear dispersion relation is required. 
For circular polarization, in which the variables $n$, $\gamma$, $|E|^2$, 
$|\pperp|^2$,  and $\ppar$ are all 
independent of phase, this is 
\eqb
\omega^2&=&\omega_{\rm p}^2+c^2k^2
\nonumber\\
&=&\gammaw^2\omega_{\rm p}^2
\label{dispersion}
\eqe
and equation~(\ref{ampere1}) reduces to 
\eqb
\frac{e^2|E|^2}{m^2c^2\omega^2}&=&
|\pperp|^2 \enspace.
\label{amperecirc}
\eqe

Equations (\ref{continuity1}) and (\ref{energy1}) can be integrated
and combined with (\ref{dispersion}) and (\ref{amperecirc}) to give 
\eqb 
J&=&r^2 2n\ppar \enspace,
\label{continuity2}\\
\mu&=&\gamma+\frac{\gammaw^2\betaw\pperp^2}{\ppar} \enspace,
\label{constK}
\eqe 
where $J$ and $\mu$ are the constants of integration. 
Equation~(\ref{constK}) simply expresses the conservation of energy,
in which the first term is the particle energy flux and the second the 
Poynting flux (per particle, in each case). In a pulsar wind,
both of these quantities, and also the particle flux $J$,
are expected to be positive, corresponding to
outward-going fluxes. Thus, the relevant parameter space is restricted to
$\ppar\ge0$, $\betaw\ge0$, and $\gamma\le\mu$, and 
the bounding line $\mu=\gamma$ corresponds to waves of vanishing amplitude,
$\pperp=0$.

As a wave packet 
propagates, its frequency remains locked to that of the pulsar. This
determines the radial dependence of the wave via the continuity equation
(\ref{continuity2}) and the dispersion relation (\ref{dispersion}): 
$\ppar\propto \gammaw^2/r^2$. In terms of the normalized radius 
$R$ defined in 
\citet[Eq.~31]{2012ApJ...745..108A}:
\eqb
R=r(\Omega_{\rm s}L\omega^2/4\pi c\dot{N}^2e^2)^{1/2},
\eqe
this relation is
\eqb
\ppar&=&\frac{\mu\gammaw^2}{R^2}\enspace.
\label{ppareq}
\eqe 
(Note that the cut-off radius $r_{\rm c}$ lies very close to, but just outside,
the point $R=1$.) 

Using $\gamma^2=1+\ppar^2+\pperp^2$ 
to eliminate $\pperp$ and equation~(\ref{ppareq}) to eliminate $\ppar$, 
we rewrite the equation of conservation
of energy flux per particle 
(\ref{constK}) as 
\eqb
\frac{\mu^2\betaw^2\gammaw^4}{R^4}+
(\mu-\gamma)\frac{\mu\betaw}{R^2}-\betaw^2(\gamma^2-1)&=&0\enspace.
\label{quadratic}
\eqe 
This equation determines the radial dependence of the group
speed of the wave $\betaw$, because, according to the entropy
equation (\ref{entropy1}), $\gamma=\,$constant. 
The radial dependence of 
$\ppar$, $\pperp$, $\left|E\right|$ and $n$ follow from (\ref{ppareq}),
(\ref{constK}), (\ref{amperecirc}) and (\ref{continuity2}).
The ram pressure,
normalized to the value of $L/\Omega_{\rm s}r^2c$ at the cut-off radius
is
\eqb
\tilde{P}&=&\frac{\nuw(R)}{\mu R^2}\enspace,
\label{rampressure}
\eqe
where $\nuw(R)$, 
which is the ratio of the momentum flux density to the particle flux
density, is defined in analogy with 
(\ref{constK}) as
\eqb
\nuw(R)&=&\ppar + \frac{\gammaw^2\pperp^2\left(1+\betaw^2\right)}{2\ppar}\enspace.
\label{constnu}
\eqe

For circular polarization, Eq.~(\ref{quadratic}) also allows one to construct
the electromagnetic Hugoniot curve found by 
\citet{2012ApJ...745..108A}. Whereas the dispersion curves
follow from the additional constraint
$\gamma=\,$constant, 
the Hugoniot curve is the locus of points under 
the constraint $\nuw(R)=\nu=\,$constant. 
In the limit of a flow carrying zero Poynting flux, which can convert
only into a wave of zero amplitude, these two curves 
coincide, $\gamma=\mu=\nuw=\,$constant, and lie on the line
\eqb
R&=&\gammaw\mu^{1/2}\left(\mu^2-1\right)^{-1/4}\enspace.
\label{weakwaves}
\eqe

\subsection{Confined and freely expanding modes}
\label{matching}

\begin{figure}
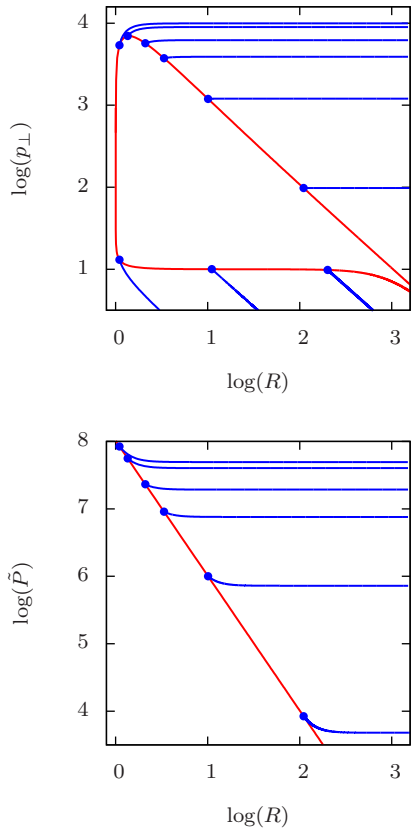

\input{fig1a.tex}
\input{fig1b.tex}
\caption{\label{fig1} 
The electromagnetic Hugoniot curve (red) for 
$\mu=10^4$, $\sigma=100$ and the 
radially propagating wave modes (blue) for circularly polarized nonlinear waves
launched at the blue dots.
Top panel: $\pperp$ 
(lower branch: free-escape mode; higher branch: confined mode).
Bottom panel: the ram pressure $\tilde{P}$, defined in Eq.~(\ref{rampressure}) 
for the confined mode only.}
\end{figure}

The radial evolution of $\pperp$ and $\tilde{P}$, for $\mu=10^4$
are shown in Fig.~\ref{fig1} for various values of $\gamma$, 
together with the Hugoniot curve for a particular choice of $\sigma$
(and, hence, $\nu$).

At large $R$, two
modes propagate: a freely expanding mode in which the second term in
(\ref{quadratic}) is negligible: 
\eqb
\pperp&\rightarrow&\sqrt{\mu\left(\mu-\gamma\right)}/R\enspace,
\\
\ppar&\rightarrow&\left(\gamma^2-1\right)^{1/2}\enspace,
\\
\betaw&\rightarrow&1\enspace,
\\
\gammaw&\rightarrow&\left(\gamma^2-1\right)^{1/4}R/\sqrt{\mu} \enspace,
\\
\tilde{P}&\rightarrow&
\frac{\mu-\gamma+\sqrt{\gamma^2-1}}{\mu R^2}\enspace,
\eqe 
and a
confined mode in which the first term in (\ref{quadratic}) is negligible:
\eqb 
\pperp&\rightarrow&\left(\gamma^2-1\right)^{1/2}\enspace,
\\
\ppar&\rightarrow&\mu/R^2\enspace,
\\
\betaw&\rightarrow&
\frac{\mu\left(\mu-\gamma\right)}{\left(\gamma^2-1\right)R^2}\enspace,
\\
\gammaw&\rightarrow&1\enspace,
\\
\tilde{P}&\rightarrow&
\frac{\gamma^2-1}{2\mu^2}\enspace.
\label{pasymptotic}
\eqe 
In the free-expansion mode, the direction
of the fluid velocity aligns itself with the radial direction as the
mode moves outwards. The radial fluid velocity tends to a constant value,
the density and ram pressure drop as $1/R^2$, and the wave amplitude decreases
as $1/R$. In the confined mode, on the other hand, the fluid velocity turns 
towards the transverse plane. The radial particle flow stagnates,
and the ram pressure, density and wave amplitude all tend to constant values,
such that the proper plasma frequency equals the pulsar rotation frequency.

\begin{figure}
\input{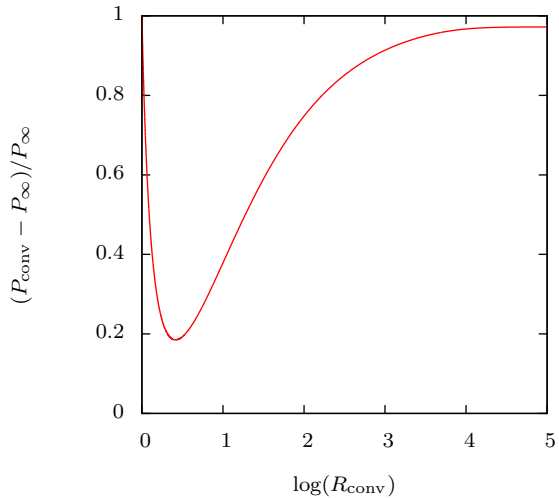}
\caption{\label{deltalogp} 
The fractional difference between the ram pressure $P_{\rm conv}$ 
at the conversion point $R_{\rm conv}$ and its value $P_\infty$ at $R\rightarrow\infty$, 
plotted for the confined mode with $\mu=10^4$ and $\sigma=100$.
}
\end{figure}

In a wind-nebula system, in which the outflow is confined by the
external medium in a slowly expanding bubble, the wind must terminate
by decelerating to nonrelativistic speed at a point where its ram
pressure roughly equals the external pressure. This is
achieved at a shock discontinuity in an MHD picture.  
From Fig.~\ref{fig1} and Eq.~(\ref{pasymptotic}), we
see that the same effect is produced if the flow converts into a
confined superluminal mode, 
which then decelerates at almost constant ram pressure.
If this mode remained stable, it could in principle accumulate and 
fill a large volume around the conversion point, which would be seen as 
a pulsar wind nebula, as originally suggested by 
\citet{1971IAUS...46..407R}.
The location of the conversion point is 
determined by the external pressure:
the higher the surrounding pressure, the closer it lies to the pulsar. As can
be seen from Fig.~\ref{deltalogp}, the ram pressure changes by less 
than a factor of 2 between the conversion point and $R\rightarrow\infty$. 
Consequently, this point is located at roughly the position 
expected for the termination shock in the MHD picture.

\subsection{Zones of stability}
\label{stabilityzones}

However, as discussed
above, once the mode has slowed down sufficiently, it is subject to strong
parametric instabilities, which ultimately dissipate the coherent oscillation,
leaving behind the relativistic electron positron plasma and residual magnetic
field that fill the pulsar wind nebula. 
In analogy with the MHD picture, we refer to the 
point at which instabilities arise in this scenario  
as the \lq\lq termination shock\rq\rq. The region 
between the point at which the confined mode is launched and the termination
shock is then an extended \lq\lq precursor\rq\rq. 

\begin{figure}
\input{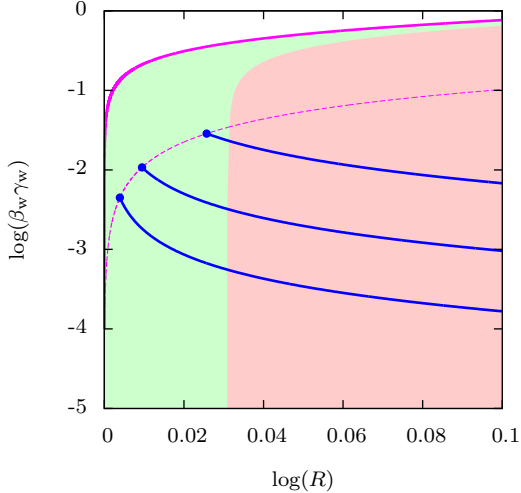}
\caption{\label{smallradius} 
The inner zone of stability. Electromagnetic Hugoniot curves are
plotted showing the group four-speed ($\betaw\gammaw$)
as a function of radius $R$ 
for $\mu =10^4$, $\sigma=100$ 
(red lines, solid: free expansion branch, dashed: confined branch). 
In blue, the radial evolution of 
three confined modes launched at different radii are plotted. 
In the green region the waves are stable according to the 
criterion given in Eq.~(\ref{stcond}). In the pink shaded region they 
are unstable. The unshaded region depicts waves with 
inwardly directed Poynting flux which are not relevant to pulsar winds. 
}
\end{figure}

The propagation and stability properties of the precursor are completely 
determined by equations (\ref{quadratic}) and (\ref{stcond}) and can be
extracted analytically and plotted relatively straightforwardly. 
In pulsars, 
we are concerned with relativistic flows of low mass-loading, $\mu\gg1$,
which can be assumed to be supersonic $\sigma\lesssim\mu^{2/3}$. Therefore,
we present here only simplified expressions obtained to lowest order in
an
expansion in powers of $\epsilon\sim\mu^{-1/3}\sim\sigma^{-1/2}$. 
In this approximation, the stability criterion (\ref{stcond})
for a stationary wave mode ($\betaw=0$) is
\eqb
S&\approx&\mu^2\left(R^2-2\sqrt{R^4-1}\right)/R^2\,>\,0\enspace.
\eqe
Thus, stationary modes, for which the H-frame coincides with the 
laboratory frame, are stable only close to the cut-off radius, in the range
$1<R<(4/3)^{1/4}$. This region is illustrated in Fig.~\ref{smallradius}.
In fact, the region of stability extends also to modes which are not stationary 
in the lab.\ frame. As is clear from the definition of $S$ in 
Eq.~(\ref{stcond}), the
waves are formally stable along the limiting line for weak waves (\ref{weakwaves}),
on which $\pperp=0$. But the region of stability reduces to a thin layer in the 
neighborhood of this line as $R$ increases. Assuming $\gammaw\sim R\gtrsim\sqrt{\sigma}$,
we find approximate roots of $S$ at
\eqb
R&=&\gammaw(8\pm 4\sqrt{3})\enspace.
\label{approxstabil}
\eqe
Thus, at large $\gammaw$, 
the zone of stability lies in the range
$\gammaw<R<1.07\gammaw$.

\begin{figure}
\input{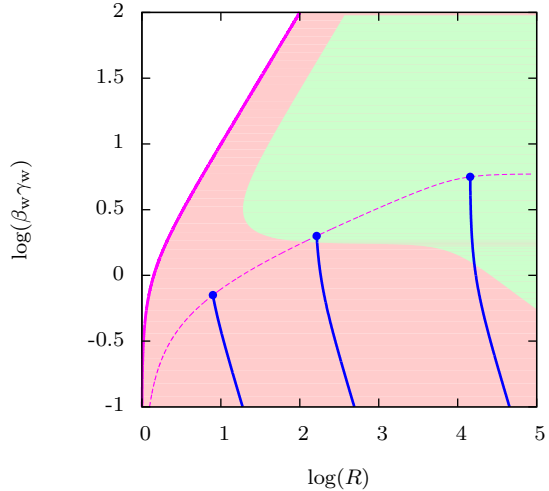}
\caption{\label{largeradius} 
The outer zone of stability. The same Hugoniot curves are plotted as in 
Fig.~\ref{smallradius}, together with three confined modes, launched at
larger radius. The pink and green shadings again depict unstable and 
stable regions, respectively. The inner zone of stability lies close to the 
Hugoniot curve of the free escape mode, and is not visible on the scale of 
this figure. 
}
\end{figure}

Equation~(\ref{approxstabil}) reveals a second stable region, which, for 
large $\gammaw$, lies at $R>14.9\gammaw$. This region is illustrated in 
Fig.~\ref{largeradius}. For $R\sim\sigma$, it is bounded from 
below by the line $\gammaw=2$, but this boundary drops 
to lower wave group speeds at larger $R$. The stable zone vanishes inside
a critical radius, which is 
roughly where the line $\gammaw=2$ intersects the line
$R=\gammaw\left(8+4\sqrt{3}\right)$, at $R=30$.

The Hugoniot curves for $\sigma=100$ are also shown in
Figs.~\ref{smallradius} and \ref{largeradius}. Expanding the
expression for the confined mode branch to lowest order in 
$\epsilon\sim\mu^{-1/3}\sim\sigma^{-1/2}$, 
and combining this with the approximate expression for
the lower bound of the stability curve 
(see Appendix~\ref{appendixapproximations}) shows that the launched waves
are stable, provided the launching radius exceeds a critical value
\eqb R&>&R_{\rm crit} \nonumber\\ &\approx& 100\enspace.
\label{criticalstab}
\eqe

The length of a precursor consisting of a stable, nonlinear
electromagnetic wave can be estimated from Figs~\ref{smallradius} and
\ref{largeradius}. In the inner stable region, the slowing down of the
wave plays a relatively minor role. Even at constant group speed, the
precursor wave enters the unstable zone after propagating for at most
4\% of the radius. Depending on the pulsar parameters, which fix the
relationship of the dimensionless scaled radius $R$ to the
light-cylinder radius, the stable precursor can nevertheless extend
over many wavelengths. In the outer stable region, however, it is the
wave deceleration that drives the precursor wave into the unstable
region. The extent in radius over which this occurs can be estimated
as $\Delta R/R\approx\diff\log R/\diff\log\left(\betaw\gammaw\right)$. Using
Eq.~(\ref{quadratic}), we find for the confined mode $\Delta
R/R\approx 1/8$, so that here again, the precursor extends over a
relatively small fraction ($\sim10\%$) of the radius, which may,
however, correspond to many wavelengths.

\section{Discussion}
\label{discussion}

The effects of spherical geometry on strong, radially propagating 
electromagnetic modes are potentially important for the structure of 
pulsar winds and their termination shocks. However, previous work on this topic 
\citep{1984A&A...139..417A,2010PPCF...52l4029K}, 
is flawed because it assumes the divergence of the radial 
momentum flux vanishes. In fact, as we show above and in Appendix~\ref{appendixentropy},
the radial momentum equation does not yield a conservation equation; 
as the modes propagate, the transverse and parallel degrees of freedom
exchange energy with each other, keeping the total energy
is conserved. We show that, for arbitrary polarization, there
exists an additional conserved quantity --- the phase-averaged Lorentz factor of the
particles $\left<\gamma\right>$ --- which, together with the conservation of
energy and particle fluxes, enables the system to be integrated.
 
Nevertheless, the case of general polarization remains unwieldy, so that 
the discussion we present of the mode properties is mainly 
restricted to circularly
polarized modes, for which phase-averages can be dropped. 
The dispersion curves, which we here call electromagnetic 
Hugoniot curves, were found by 
\citet{2012ApJ...745..108A} for linearly polarized modes with
vanishing phase-averaged magnetic field. They closely resemble those for
circular polarization, suggesting that this simplification is not an important
restriction, at least in the region close to the equatorial plane. 

The electromagnetic Hugoniot curve admits two solutions 
for radial propagation \citep{2012ApJ...745..108A}. By examining
the asymptotic behavior of waves launched on these branches, we
identify them as a free-escape and a confined mode, with very different
properties. The group speed of the former accelerates, whereas that of
the latter decelerates during outward propagation from the launching
point. We argue that, in outflows such as those from pulsars surrounded
by nebulae, only the confined mode can provide a self-consistent
solution matching a relativistic, highly magnetized wind to a
slowly expanding, weakly magnetized nebula. We show that the external pressure
determines the radius at which a confined mode must be launched in
order to realize such a solution. The higher the pressure, the closer to the 
star the launching, or conversion, takes place. However, we do not discuss the 
physical mechanism by which the wind converts from one
mode to the other. This question demands a different approach, and most likely 
necessitates either 
two-fluid or particle-in-cell simulations. It has been addressed recently 
\citep{amanokirk13}, but only for a very limited range of physical
parameters, corresponding to a launching radius close to the critical
radius $r_{\rm c}$.

\tableone

A confined mode becomes unstable against longitudinal density
perturbations when the particle streaming through the wave drops below
a critical value. For plane, circularly polarized waves, we adopt a
simple stability condition (\ref{stcond}) against long-wavelength,
parallel fluctuations proposed by \citet{1978JPlPh..20..313L}.  In
fact, obliquely propagating fluctuations may change this simple
condition if they turn out to be more unstable than the parallel ones
\citep{1978JPlPh..20..313L}, but we are not aware of a suitable
alternative criterion.  On the other hand, waves which have
a nonvanishing component of the transverse magnetic field, i.e., those
which are launched at higher latitudes in pulsar winds, may be
stabilized by the presence of this field \citep{1980PhRvA..22.1293A}.

Using this criterion, we find two regions in which confined modes are
stable when launched, although they subsequently propagate into an
unstable region.  In terms of the normalized radius, these are an
outer zone, $R\gtrsim100$, and an inner zone $R\approx1$. These zones
and the location of their boundaries are essentially independent of
the parameters $\mu$ and $\sigma$ of the pulsar wind, provided it is
relativistic, $\mu\gg1$, and supermagnetosonic,
$\sigma<\mu^{2/3}$. However, their relevance depends on where the
pressure confining a particular pulsar wind places the launching
point. This, in turn, depends on the radius normalization, which
varies from pulsar to pulsar. In terms of the light-cylinder radius
$r_{\rm L}$, the normalized radius is $R=(r/r_{\rm L})(\mu/a_{\rm L}\approx r_{\rm L}/r_{\rm c})$, where the strength parameter
at the light cylinder $a_{\rm L}$ is a dimensionless measure of the
spin-down power per unit solid angle: $a_{\rm L}=3.4\times
10^{10}\left[\left(L/10^{38}\textrm{erg\,s}^{-1}\right)
  \left(4\pi/\Omega_{\rm s}\right)\right]^{1/2}$. 

We show in Fig.~\ref{deltalogp}
that the pressure at the conversion point lies close to $L/\Omega_{\rm s}r^2c$,
which can be written as
\eqb
P_{\rm conv}&=&\frac{10^{-6}}{R^2}\left(\frac{\mu}{10^4}\right)^2
\left(\frac{P_{\rm pulsar}}{1\,\textrm{second}}\right)^{-2}
\ \textrm{dyn\,cm}^{-2} \enspace,
\eqe
where $P_{\rm pulsar}$ is the rotation period of the pulsar.
This must be compared with the nebular pressure, 
usually estimated from the volume-averaged,
equipartition magnetic and particle energy densities required to
produce the nebular synchrotron emission
\citep{1984ApJ...283..694K}. 
These estimates, together with the implied normalized radius of the conversion point $R_{\rm conv}$
are listed in Table~\ref{pwntable} for 
those pulsar wind nebulae
studied by 
\citet{2011MNRAS.410..381B}.
We see from this table that the isolated pulsars can be expected to 
launch stable precursor waves at the termination shock, since 
in all but one case $R_{\rm conv}>100$. The only exception is the pulsar in the supernova remnant W44, which, however, is known to be interacting with a dense molecular cloud \citep{1999ApJ...524..179C}.  

The PWNe of isolated pulsars in starburst galaxies have recently been
proposed as sources
of the very high energy emission from those galaxies 
\citep{2012APh....35..797M,2013MNRAS.429L..70O}. 
The pressure in the 
interstellar medium in these objects ($\sim10^{-9}\,\textrm{dyn\,cm}^{-2}$) 
is some three orders of magnitude
greater than in the Milky Way. If this scaling is reflected in the
pressure in PWN bubbles, the estimated conversion radii 
fall by a factor of 30. From Table~\ref{pwntable}, we see that 
the precursor waves would in this case be launched in the unstable
zone $R<100$. It seems reasonable to suppose that the shock structure
will be strongly influenced by such a change. However, the possible 
implications for particle acceleration and the resulting 
non-thermal photon emission remain unknown. 

The inner zone of stability, on the other hand, may be of importance
for pulsar winds in high pressure environments, such as
outflows from a companion star.
Several gamma-ray binaries are likely to fall into this class 
\citep{2006A&A...456..801D}
but so far only one of them, B1259$-$63, harbors a pulsar
with a measured period. 
Its spin-down power $L=8\times10^{35}\,\textrm{erg\,s}^{-1}$, 
implies $a_{\rm L}=3\times10^9$, an order of magnitude 
lower than that of the Crab. Assuming young and middle-aged pulsars have 
a pair multiplicity $\kappa=a_{\rm L}/(4\mu)$ that is at least $10^5$   
\citep{2011MNRAS.410..381B}, we find, for
B1259$-$63, $\mu\gtrsim7.6\times10^3$. The confining 
pressure at periastron, and, hence the pressure at the conversion point,
can be estimated as
$P_{\rm conv}\geq2.5\times10^{-4}\textrm{ dyn\,cm}^{-2}$ 
\citep{2006A&A...456..801D}. This implies $R_{\rm conv}\approx1$, i.e.,
the conversion point lies roughly at the critical radius inside which superluminal waves cannot
propagate. Since the confining pressure varies over the binary orbit,
it appears possible that the pulsar wind in this object will terminate
at $r<r_{\rm c}$ close to periastron, but at $r>r_{\rm c}$ at
larger binary separations. At the transition points, the conversion
point passes through the inner stable zone, and, according to 
Fig~\ref{smallradius}, a stable precursor wave is possible for a 
brief interval of phase.

\section{Conclusions}
\label{conclusions}

Our main results are  
the derivation for strong superluminal waves of an analogue of the 
hydrodynamical entropy equation, the demonstration that two 
kinds of wave, confined and freely expanding, propagate in spherical geometry, 
and the identification of two zones of stability of the confined mode
in pulsar winds. Strictly speaking, the latter two results apply only to 
circularly polarized modes, but the similarity of the dispersion curves for
linear and circular modes suggest this is not a serious restriction. 

The application of these results to PWN around isolated pulsars and around
pulsars in gamma-ray binaries suggests different termination shock structures
depending on the confining pressure. However, the implications for particle 
acceleration and, hence possible observational signatures remain a topic for
future work.

\acknowledgments{We thank Ioanna Arka, J\'er\^ome P\'etri
and Takanobu Amano for fruitful discussions.}

\appendix

\section{Radial propagation of large-amplitude waves}
\label{app1}

To solve a coupled system of two-fluid and Maxwell equations in spherical geometry, we use a standard perturbation technique, following \citet{1984A&A...139..417A} and \citet{2011ApJ...729..104K,2011ApJ...736..165K}. We assume two timescales, on which the wave properties change: a fast scale -- determined by the pulsar rotation period, and a slow scale, on which the wave quantities evolve slowly due to spherical expansion. The \lq\lq fast\rq\rq\ variable is the WKB-like phase of a wave 
\eqb 
\phi=\omega\left(t-\int^r\frac{dr'}{c\beta(r')}\right)\enspace,
\label{wkbphase}
\eqe
where $\beta=1/\betaw$ is the superluminal phase velocity of a wave, $\betaw$ is the subluminal group speed. 
The \lq\lq slow\rq\rq\ radial coordinate is defined by the light cylinder distance $r_{\rm L}$ as
\eqb \rho=\epsilon \frac{r}{r_{\rm L}}\enspace, \eqe
where $\rho\sim1$ and $\epsilon\sim r_{\rm L}/r\ll1$ is a small parameter. 
The time and space derivatives, expressed in terms of the new variables $\phi$ and $\rho$, take the following form: 
\eqb 
\der{t} \rightarrow \omega\der{\phi}\enspace, \enspace
\der{r} \rightarrow \epsilon\frac{\omega}{c}\der{\rho}-\frac{\omega\betaw}{c}\der{\phi}\enspace, \enspace
\gamma\derd{t} \rightarrow \epsilon\frac{\omega}{c} p_{\parallel}\der{\rho}
+
\omega\Delta\der{\phi}\enspace, 
\label{derivfull}
\eqe
where $\Delta=\gamma-\betaw\ppar$. Expressions (\ref{derivfull}) are substituted into the equations that govern wave propagation (e.g., \citealt{1974JPlPh..12..297C,1976JPlPh..15..335K}), and, in addition, all the dependent variables in these equations are expanded in $\epsilon$ to the first order, i.e., $\gamma=\gamma^{(0)}+\epsilon\gamma^{(1)}$ etc. The proper density of each plasma species and the electromagnetic fields are expressed in a dimensionless form: $n\rightarrow n m\omega^2/8\pi e^2$, $E\rightarrow eE/mc\omega$, $B\rightarrow eB/mc\omega$, where  $E=E_y+iE_z$ and $B=B_y+iB_z$ are complex quantities (so are the transverse momenta $\pperp=p_y+ip_z$). 

In the lowest order in $\epsilon$ we obtain the equations describing plane waves. Those are the continuity equation:
\eqb
\frac{\partial}{\partial\phi}\left(\nexpand[0]\Deltaexpand[0]\right)
=0 
\enspace,
\label{contzero} 
\eqe
Faraday's and Amp\`ere's laws:
\begin{align}
\betaw\frac{\partial \Eexpand[0]}{\partial\phi}
+i\frac{\partial \Bexpand[0]}{\partial\phi}
&=0\enspace, \label{faradayzero} \\
-\betaw\frac{\partial \Bexpand[0]}{\partial\phi}
+i\frac{\partial \Eexpand[0]}{\partial\phi}+i\nexpand[0]\pperpexpand[0]
&=0 
\enspace,
\label{Amperezero} 
\end{align}
and momentum/energy conservation multiplied by $n$:
\begin{align}
\nexpand[0]\Deltaexpand[0]\frac{\partial \pparexpand[0]}{\partial\phi}
+\nexpand[0]\textrm{Im}\left(\pperpexpand[0] \Bexpand[0]^*\right)
&=0 \enspace,
\label{pparzero}
\\
\nexpand[0]\Deltaexpand[0]\frac{\partial \pperpexpand[0]}{\partial\phi}
-\nexpand[0]\left(\gammaexpand[0] \Eexpand[0]+i\pparexpand[0] \Bexpand[0]\right)
&=0 \enspace,
\label{pperpzero}
\\
\nexpand[0]\Deltaexpand[0]\frac{\partial \gammaexpand[0]}{\partial\phi}
-\nexpand[0]\textrm{Re}\left(\pperpexpand[0] \Eexpand[0]^*\right)
&=0
\enspace.
\label{gammazero}
\end{align}
Equations (\ref{faradayzero}) and (\ref{Amperezero}) imply $\Bexpand[0]=i\betaw\Eexpand[0]$ and 
$\nexpand[0]\pperpexpand[0]=-(1-\betaw^2)\partial\Eexpand[0]/\partial\phi$. Using these relations in equations (\ref{pparzero}) and (\ref{gammazero}), we arrive at the 
result that the quantity $\deltaexpand[0]=\pparexpand[0]-\betaw\gammaexpand[0]$ is phase independent.

In the first order in $\epsilon$ the equation 
of continuity reads: 
\eqb
\frac{\partial}{\partial\phi}\left(
\nexpand[0]\Deltaexpand[1]+
\nexpand[1]\Deltaexpand[0]
\right)
+\frac{1}{{\rho}^2}\frac{\partial}{\partial {\rho}}\left({\rho}^2\nexpand[0]\pparexpand[0]\right)
=0 
\label{contone} 
\enspace,
\eqe
Faraday's and Amp\`ere's laws are:
\begin{align}
-\betaw\frac{\partial \Eexpand[1]}{\partial\phi}
-i\frac{\partial \Bexpand[1]}{\partial\phi}
+
\frac{1}{{\rho}}\frac{\partial}{\partial {\rho}}\left({\rho}\Eexpand[0]\right)&=0 \enspace,
\label{appfaraday} \\
-\betaw\frac{\partial \Bexpand[1]}{\partial\phi}
+i\frac{\partial \Eexpand[1]}{\partial\phi}
+i\left(\nexpand[0]\pperpexpand[1]+\nexpand[1]\pperpexpand[0]\right)
+\frac{1}{{\rho}}\frac{\partial}{\partial {\rho}}({\rho}\Bexpand[0])&=0 
\enspace,
\label{appAmpere} 
\end{align}
and momentum/energy equations, after multiplying by $n$, give:
\begin{align}
\left(\nexpand[1]\Deltaexpand[0]+\nexpand[0]\Deltaexpand[1]\right)\frac{\partial \pparexpand[0]}{\partial\phi}+\nexpand[0]\Deltaexpand[0]\frac{\partial \pparexpand[1]}{\partial\phi}
+\left[n\,\textrm{Im}\left(\pperp B^*\right)\right]^{(1)}
+\nexpand[0]\pparexpand[0]\frac{\partial \pparexpand[0]}{\partial {\rho}} 
&=\nexpand[0]\frac{|\pperpexpand[0]|^2}{\rho} \enspace,
\label{pparone}
\\
\left(\nexpand[1]\Deltaexpand[0]+\nexpand[0]\Deltaexpand[1]\right)\frac{\partial \pperpexpand[0]}{\partial\phi}
+\nexpand[0]\Deltaexpand[0]\frac{\partial \pperpexpand[1]}{\partial\phi}
-\left[n\left(\gamma E+i\ppar B\right)\right]^{(1)}
+\nexpand[0]\pparexpand[0]\frac{\partial \pperpexpand[0]}{\partial {\rho}} 
&=-\nexpand[0]\frac{\pparexpand[0]\pperpexpand[0]}{\rho} \enspace,
\label{pperpone}
\\
\left(\nexpand[1]\Deltaexpand[0]+\nexpand[0]\Deltaexpand[1]\right)\frac{\partial \gammaexpand[0]}{\partial\phi}
+\nexpand[0]\Deltaexpand[0]\frac{\partial \gammaexpand[1]}{\partial\phi}
-\left[n\,\textrm{Re}\left(
\pperp E^*\right)\right]^{(1)}
+\nexpand[0]\pparexpand[0]\frac{\partial \gammaexpand[0]}{\partial {\rho}} 
&=0
\enspace.
\label{gammaone}
\end{align}

To ensure that the first-order quantities have regular (nonsecular) behavior, we impose the condition that they are periodic in $\phi$. This suffices to determine the slow radial dependence of the lowest-order, phase-averaged terms. For example, integrating Eq.~(\ref{contone}) over $\phi$, we obtain
\eqb  \derdrho\left(\rho^2\av{\nexpand[0]\pparexpand[0]}\right)=0 \enspace. \eqe
Next, we express the transverse momentum in terms of the fields using Faraday's and Amp\`ere's laws, (\ref{appfaraday}) and (\ref{appAmpere}), which allows us to calculate the phase-averaged forces:
\begin{align}
\av{\left[n\,\textrm{Re}\left(\pperp E^*\right)\right]^{(1)}}
&=-\frac{1}{\rho^2}\frac{\partial}{\partial\rho}\av{\rho^2\betaw|\Eexpand[0]|^2} \enspace, \label{forceelectric}\\
\av{\left[n\,\textrm{Im}\left(\pperp B^*\right)\right]^{(1)}}
&=\frac{1}{2}\frac{1}{\rho^2}\frac{\partial}{\partial\rho}\av{\rho^2\left(1+\betaw^2\right)|\Eexpand[0]|^2} \enspace. \label{forcelorentz}
\end{align}
Substituting those into the equations of motion (\ref{gammaone}) and (\ref{pparone}), we obtain 
the conservation of the total energy flux and evolution of the radial momentum flux, respectively: 
\begin{align}
\derdrho\left[\rho^2\av{\nexpand[0]\pparexpand[0]
\gammaexpand[0]+\betaw\left|\Eexpand[0]\right|^2}\right]&=0\enspace, \\
\derdrho\left[\rho^2\av{\nexpand[0]\pparexpand[0]^2
+\frac{1}{2}\left(1+\betaw^2\right)\left|\Eexpand[0]\right|^2}\right]&=\frac{1}{\rho}\av{\nexpand[0]\pperpexpand[0]^2}\enspace.
\end{align}

\section{Conservation of $\av{\gamma}$}
\label{appendixentropy}

The analogue of the hydrodynamic entropy equation is obtained by multiplying (\ref{gammaone}) by $\betaw$ and subtracting it from (\ref{pparone}):
\begin{multline}
\left(\nexpand[1]\Deltaexpand[0]+\nexpand[0]\Deltaexpand[1]\right)\frac{\partial \deltaexpand[0]}{\partial\phi}
+\nexpand[0]\Deltaexpand[0]\frac{\partial \deltaexpand[1]}{\partial\phi}
+\betaw\left[n\,\textrm{Re}\left(
\pperp E^*\right)\right]^{(1)}
+\left[n\,\textrm{Im}\left(
\pperp B^*\right)\right]^{(1)} \\
+\nexpand[0]\pparexpand[0]\frac{\partial \pparexpand[0]}{\partial {\rho}}
-\betaw\nexpand[0]\pparexpand[0]\frac{\partial \gammaexpand[0]}{\partial {\rho}}
-\nexpand[0]\frac{|\pperpexpand[0]|^2}{\rho}
=0 \enspace.
\label{entropyone}
\end{multline}
It is shown in Appendix~\ref{app1} that the lowest order equations imply that $\delta^{(0)}$ is phase independent, and the first order quantities are required to be periodic in $\phi$. Therefore, phase averaging of (\ref{entropyone}) immediately cancels first two terms, and the remaining part takes the form
\begin{multline}
-\betaw\frac{1}{\rho^2}\frac{\partial}{\partial\rho}\av{\rho^2\betaw|\Eexpand[0]|^2}
+\frac{1}{2}\frac{1}{\rho^2}\frac{\partial}{\partial\rho}\av{\rho^2\left(1+\betaw^2\right)|\Eexpand[0]|^2} \\
+\av{\nexpand[0]\pparexpand[0]\frac{\partial \pparexpand[0]}{\partial {\rho}}} 
-\betaw\av{\nexpand[0]\pparexpand[0]\frac{\partial \gammaexpand[0]}{\partial {\rho}}} 
-\frac{1}{\rho}\av{\nexpand[0]|\pperpexpand[0]|^2}
=0\enspace,
\label{entropytwo}
\end{multline}
where we have used the expressions for phase-averaged forces (\ref{forceelectric}) and (\ref{forcelorentz}). In the following we omit superscript \lq\lq 0\rq\rq\ and further simplify the equation (\ref{entropytwo}) by the momentum relation $\ppar^2=\gamma^2-\pperp^2-1$, so that it reads 
\eqb 
\frac{\av{E^2}}{\gammaw^2\rho}+\frac{1}{2\gammaw^2}\frac{\partial\av{E^2}}{\partial\rho}+n_0\Delta_0\frac{\partial\av{\gamma}}{\partial\rho}-\frac{1}{2}\av{n\frac{\partial\pperp^2}{\partial\rho}}-\frac{\av{n\pperp^2}}{\rho}=0 \enspace.
\label{entropy}
\eqe
The first term in (\ref{entropy}) cancels the last term, since:   
\eqb
\av{E^2}=\int_0^{2\pi} d\phi E^2=\int_0^{2\pi} d\phi \frac{\partial\pperp^*}{\partial\phi}\frac{\partial\pperp}{\partial\phi}=-\int_0^{2\pi} d\phi\pperp^*\frac{\partial^2\pperp}{\partial\phi^2}=\int_0^{2\pi} d\phi \gammaw^2n\pperp^2=\gammaw^2\av{n\pperp^2} \enspace.
\label{avEsq}
\eqe
Here in the second step we use the equation (\ref{pperpzero}) and (\ref{faradayzero}), which together imply $\Eexpand[0]=\partial\pperpexpand[0]/\partial\phi$, and we integrate once by parts.  Further simplification follows from Eq.~(\ref{Amperezero}): $\partial^2\pperp/\partial\phi^2=\partial\Eexpand[0]/\partial\phi=-\gammaw^2n\pperp$. 
The same relations prove that the second term in the entropy equation (\ref{entropy}) cancels the fourth term: 
\begin{multline}
\av{n\frac{\partial\pperp^2}{\partial\rho}}=\int_0^{2\pi} d\phi \left( n\pperp\frac{\partial\pperp^*}{\partial\rho}+n\pperp^*\frac{\partial\pperp}{\partial\rho} \right)=-\frac{1}{\gammaw^2}\int_0^{2\pi} d\phi \left( \frac{\partial E}{\partial\phi}\frac{\partial\pperp^*}{\partial\rho}+\frac{\partial E^*}{\partial\phi}\frac{\partial\pperp}{\partial\rho} \right) 
\\=
\frac{1}{\gammaw^2}\int_0^{2\pi} d\phi \left( E\frac{\partial E^*}{\partial\rho}+E^*\frac{\partial E}{\partial\rho} \right)=\frac{1}{\gammaw^2}\frac{\partial}{\partial\rho}\av{E^2} \enspace.
\end{multline}
Thus, Eq.~(\ref{entropy}) implies that independently of the wave polarization the phase-averaged Lorentz factor of the particles stays constant during the radial expansion: 
\eqb \frac{\partial\av{\gamma}}{\partial\rho}=0 \enspace. \eqe

\section{Linear polarization}
\label{linear}

For linearly polarized waves all the quantities are phase dependent. It is convenient to use the phase variable $y=E/E_0$, chosen to be $y=1$ for the phase $\phi$, for which the electric field takes the maximum value $E_0$. 
The nonlinear dispersion relation follows from the periodicity requirement
\eqb 
1&=&\frac{2}{\pi}\int_0^{1}\frac{dy}{|dy/d\phi|} 
\enspace, 
\label{lindispersion}
\eqe
and Amp\`ere's law takes the form
\eqb \frac{dy}{d\phi}=-\frac{\omega_{\rm p}^2\gammaw^2\pperp}{E_0\omega^2}=-\frac{\Delta_0\omega_{\rm p0}^2\gammaw^2}{E_0\omega^2}\frac{\pperp}{\Delta} \enspace, \eqe 
where $\omega_{\rm p0}^2=8\pi e^2n_0/m$, and the continuity equation (\ref{contzero}) implies that $n\Delta\equiv n_0\Delta_0$ is phase-independent. 
In radial evolution equations all the phase averages have to be calculated explicitly, i.e.,
\eqb 
\av{X}&=&\frac{2}{\pi}\int_0^{1}X(y)\frac{dy}{|dy/d\phi|} \enspace.
\label{averaging}
\eqe
Dependence of plane-wave particle momenta $\gamma$, $\pperp$, $\ppar$ on $y$ one obtains by integrating equations of motion, e.g.,  \citet{1976JPlPh..15..335K}:
\begin{align}
\gamma&=\gamma_0+\frac{\omega^2E_0^2}{2\omega_{\rm p0}^2\Delta_0\gammaw^2}(1-y^2)\enspace, \label{lingamma} \\
\ppar&=\pparb+\frac{\omega^2E_0^2\betaw}{2\omega_{\rm p0}^2\Delta_0\gammaw^2}(1-y^2)\enspace, \label{linppar} \\
\pperp&=(\gamma^2-\ppar^2-1)^{1/2}\enspace, \label{linpperp}
\end{align}
where $\gamma_0^2=\pparb^2+1$. 
The set of equations which describes radial propagation of a linearly polarized wave with variables $n_0$, $\pparb$, $E_0$, $\pperp$, $\gamma_0$, $\betaw$ is given by equations (\ref{continuity1}), (\ref{energy1}), 
and (\ref{entropy1}), in which the phase averages have to be calculated explicitly according to  Eq.~(\ref{averaging}), together with (\ref{lindispersion}) and (\ref{linpperp}). 

\section{The outer stability zone}
\label{appendixapproximations}

The Hugoniot curves are determined by Eqs.~(\ref{constK}) and (\ref{constnu}), where $\nu_{\rm w}(R)=\nu$ is a constant, related to $\mu$ and $\sigma$ by (\ref{approxnu}). These equations are complemented by (\ref{ppareq}) and momentum relation 
$\gamma=(1+\pperp^2+\ppar^2)^{1/2}$. Finding $\pperp$ from (\ref{constnu}) and substituting to (\ref{constK}), we obtain a biquadratic equation for radius $R$ as a function of the wave Lorentz factor $\gammaw$. For the confined mode branch the larger root is relevant. The largest term in the expansion in $\epsilon\sim\mu^{-1/3}\sim\sigma^{-1/2}$ reads:
\eqb
R^2\approx 
\frac{\gammaw^2-2\gammaw^2 \left(2 \gammaw^2-1\right)\left(\gammaw^2-1\right)^{1/2}\left[2 \gammaw
   \left(\gammaw-\sqrt{\gammaw^2-1}\right)-1\right]^{1/2}
   -2\gammaw^3\left(2 \gammaw^2-1 \right)
   \left(\gammaw-\sqrt{\gammaw^2-1}\right)}{4\gammaw \left(2 \gammaw- \sqrt{\gammaw^2-1}\right)  -8  \gammaw^3 \left(\gammaw-\sqrt{\gammaw^2-1}\right)-1}
\enspace.
\label{expandR}
\eqe
The lower bound on $\gammaw$, and thus $R$, for which the Hugoniot curve implies stable initial conditions for a wave at launch, is obtained from the stability condition (\ref{stcond}), in which the particle momenta are transformed from the H-frame to the lab frame: $\gamma'=\gammaw(\gamma-\betaw\ppar)$, $\ppar'=\gammaw(\ppar-\betaw\gamma)$. Expanding in $\epsilon\sim\mu^{-1/3}\sim\sigma^{-1/2}$, and assuming for large radii $R\sim\epsilon^{-1}$, in the lowest order Eq.~(\ref{stcond}) takes the form 
\eqb 
\frac{2 \left(-2 \gammaw^3+\gammaw^4\right)}{\left(-1+2 \gammaw^2\right)}\gtrsim0 \enspace.
\eqe
This implies the lower bound $\gammaw\approx2$, and thus Eq.~(\ref{expandR}) implies the relation (\ref{criticalstab}). 


\end{document}